\documentstyle[editedbook,epsfig,psfig,epsf]{mq}

\def\etal{{\em et al.} }

\def\msun{$M_{\odot}$}

\def\msunyr{M$_{\odot}$ yr$^{-1}$ }

\def\ergs{erg s$^{-1}$ }

\def\cm2{cm$^2$ }
\def\se1{s$^{-1}$ }

\def\ss{SS~433}

\begin{opening}
\title{SS~433: Radio/X-ray anti-correlation and fast-time variability}
\author{Samar Safi-Harb$^{1,2}$ \& Taro Kotani $^3$}
\institute{$^1$ Physics and Astronomy, University of Manitoba, Winnipeg,
MB R3T 2N2 Canada.\\
$^2$ NSERC UFA fellow; samar@physics.umanitoba.ca\\
$^3$ NASA/Goddard Space Flight Center.}
\end{opening}

\runningtitle{SS~433}
\runningauthor{Safi-Harb \& Kotani}

\begin{document}
\vspace{-0.5cm}
\begin{abstract}
{\small
We briefly review the Galactic microquasar SS~433/W50 and
present a new RXTE spectral and timing study.
We show that the X-ray flux decreases during radio flares,
a behavior seen in other microquasars. We also
find short time-scale variability unveiling
emission regions from within the binary system.
}
\end{abstract}

\section{SS433/W50 Review}
\ss\ is the famous galactic microquasar known by its
two-sided precessing mildly relativistic (0.26c) jets
(Margon 1984). The question whether the compact object is a neutron star
or a black hole remains unanswered, in spite of its
discovery over 2 decades ago.
ASCA observations revealed Doppler blue- and red-shifted emission
lines (Kotani et al. 1996) indicating a jets' scale
of $\sim$ 10$^{13}$~cm.
Recent Chandra observations with the gratings (Marshall et al. 2002)
gave  a mass outflow rate of 
1.5$\times$10$^{-7}$ \msunyr and a kinetic power of
3.2$\times$10$^{38}$ \ergs  (at an assumed distance of 4.85 kpc).
The latter values are small compared to previous estimates
of $\sim$ 10$^{-6}$ \msunyr and 10$^{39}$-10$^{41}$
\ergs (possibly because \ss\ was in a low state during the
Chandra observations). 
\par
\ss\ is near the center of the supernova remnant (SNR) W50.
Observations in the radio (Dubner et al. 1998), millimeter
(Durouchoux \etal 2000; Sood \etal 2002 in preparation),
 infrared (Fuchs, Koch-Miramond, \& Abraham 
2001), optical (Mazeh \etal 1983),
and X-rays (Safi-Harb \& Petre 1999 and refs therein)
show that the \ss\ jets are interacting 
with the surounding inhomogeneous medium, causing the unusual
morphology and the X-ray lobes of W50.

\par
\ss\ was observed by the RXTE at several 
occasions. Here we present the preliminary results of our spectral and timing study.
A more detailed analysis and interpretation of our results will
be presented elsewhere. We also refer the reader to a companion paper
(Kotani \etal this volume) highlighting the multiwavelength 2001 campaign,
and to several other \ss\ papers in 
this volume highlighting the most recent multi-wavelength observations:
Namiki \etal
for recent Chandra observations, Fabrika \etal for the optical observations,
Fuchs \etal for the infrared observations suggesting a Wolf-Rayet star origin,
Blundell \etal and Paragi \etal for the evidence of
an equatorial outflow, Migliari
\etal for thermal reheating in the jet,
Trushkin \etal for the 6-day modulation in the quite radio emission,
and Chakrabarti \etal for the theoretical implications.

\section{Radio and X-ray anti-correlation}
RXTE observed \ss\ in 1996 ($\phi$=0.83--1.12; $\psi$=0.78--0.81),
1998 ($\phi$=0.01--1.0; $\psi$=0.1--0.2), and 2001 
($\phi$=0.1--1.3; $\psi$=0.29--0.38); where $\phi$ and $\psi$ are
the optical and precession phases, respectively.
A radio flare occurred mid March 1998 (around MJD 50890)
and in November 2001 (MJD 52215 and 52235), as shown
in Fig.~1 by the arrows. Luckily the 1998 flare 
(detected with the NRAO GBI) was
covered by two short RXTE observations, and the November 2001 flare
detected with the RATAN 600-m telescope (Kotani \& Trushkin 2001)
was monitored by  RXTE for 2 weeks (except for
November 18, 2001).

\begin{figure}[tbh]
\centering
\epsfig{file=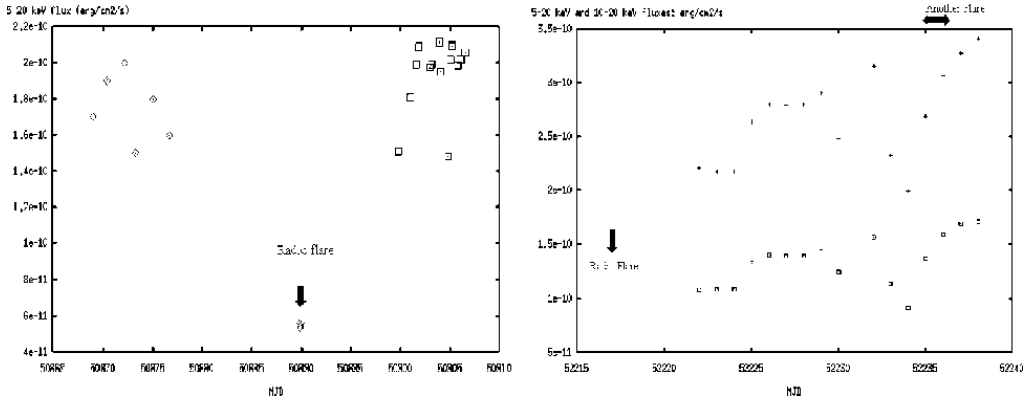, width=13.5cm}
\caption{The hard X-ray PCA fluxes of \ss\ in 1998 and 2001.}
\label{myfig}
\end{figure}

\begin{figure}[tbh]
\centering
\epsfig{file=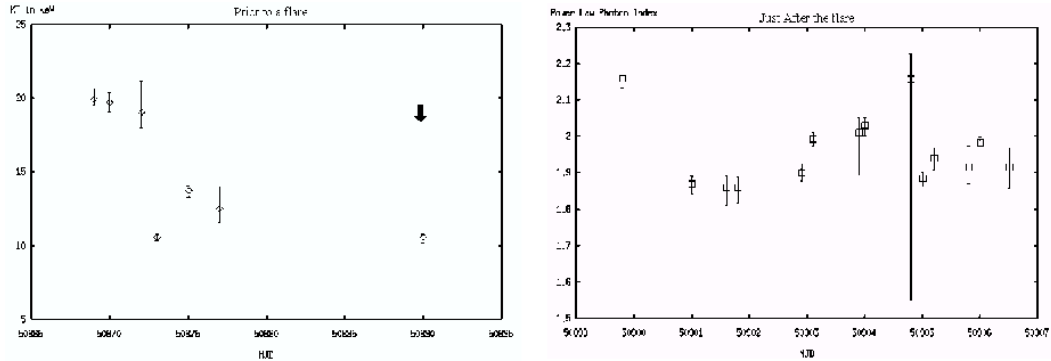, width=14cm}
\caption{The thermal bremsstrahlung temperature 
($kT$) and power law photon index ($\Gamma$) for
the 1998 observations. The arrow in the left panel shows the
radio flare (MJD 50890).}
\label{myfig}
\end{figure}


The PCA spectra are well fitted with a thermal bremsstrahlung model
(or a power law model at occasions, see below),
modified by interstellar absorption,
and with two Fe-lines: A broad line and a narrow line.
In Fig.~1, we show the hard X-ray flux variations
and its anti-correlation with the radio flares that occurred
in 1998 (left) and 2001 (right).
In Fig.~2, we show the variations of the spectral parameters,
$kT$ and $\Gamma$,
fixing $N_H$ at 0.7$\times$10$^{22}$~cm$^{-2}$.
We find that 1) the X-ray flux decreases during
the flare, 2) the spectrum softens at the onset of the flare
then hardens shortly after, and 3) the spectrum switches from thermal to power law right
after the flare in 1998.
The narrow line variations (believed to originate from
the jets) are also consistent with the redshifts
or blueshifts determined from optical spectroscopy  (Kotani \etal 2002).

\section{Fast time variability}
Previous observations revealed no variability on time-scales
shorter than 300~s. A search for variability in the 0.5-300~s range
in the ROSAT data (Safi-Harb 1997, see Fig.~3, left) using the method of
autocorrelation function 
shows flickering around 3--10~s. However, this variability does not
appear consistently in all the observation segments, and the time
scale varies from an observation segment to the other. 
Using the PCA Nov. 2001 data, we
also found fast time variability
(Fig.3, right), which we confirmed
using previous archival PCA observations. The time scale of 50--100~s
indicates a length scale $\leq$10$^{12}$cm.
Fast time variability is expected when the source enters a 
highly non-stationary regime expected from super-critical accretion
into the surface of a neutron star. 

\begin{figure}[tbh]
\hspace{-0.2cm}
\centering{
\begin{tabular}{ll}
\hspace{-0.5cm}
\epsfig{file=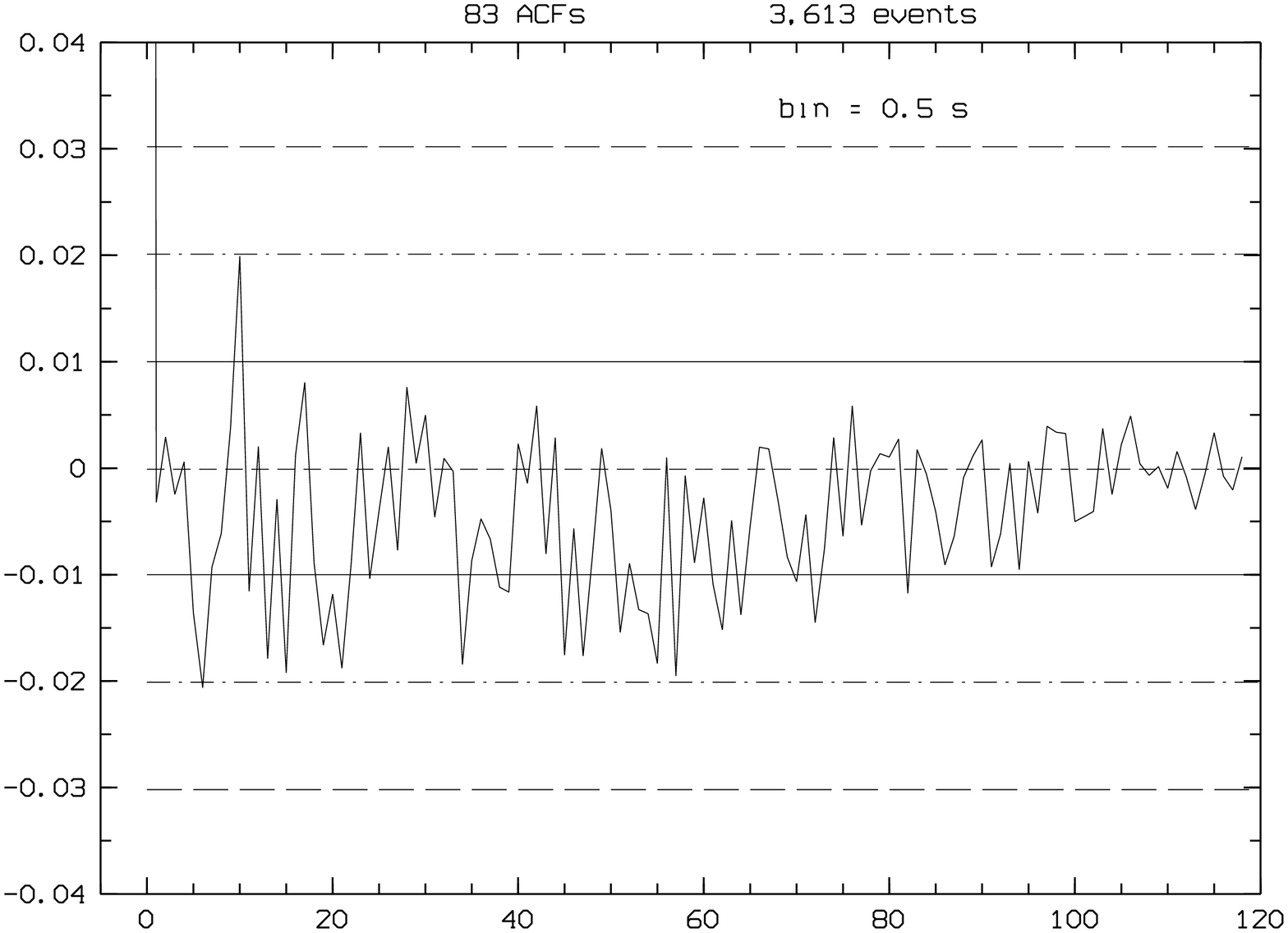,width=4.5cm, angle=270}&
\hspace{-0.5cm}
\epsfig{file=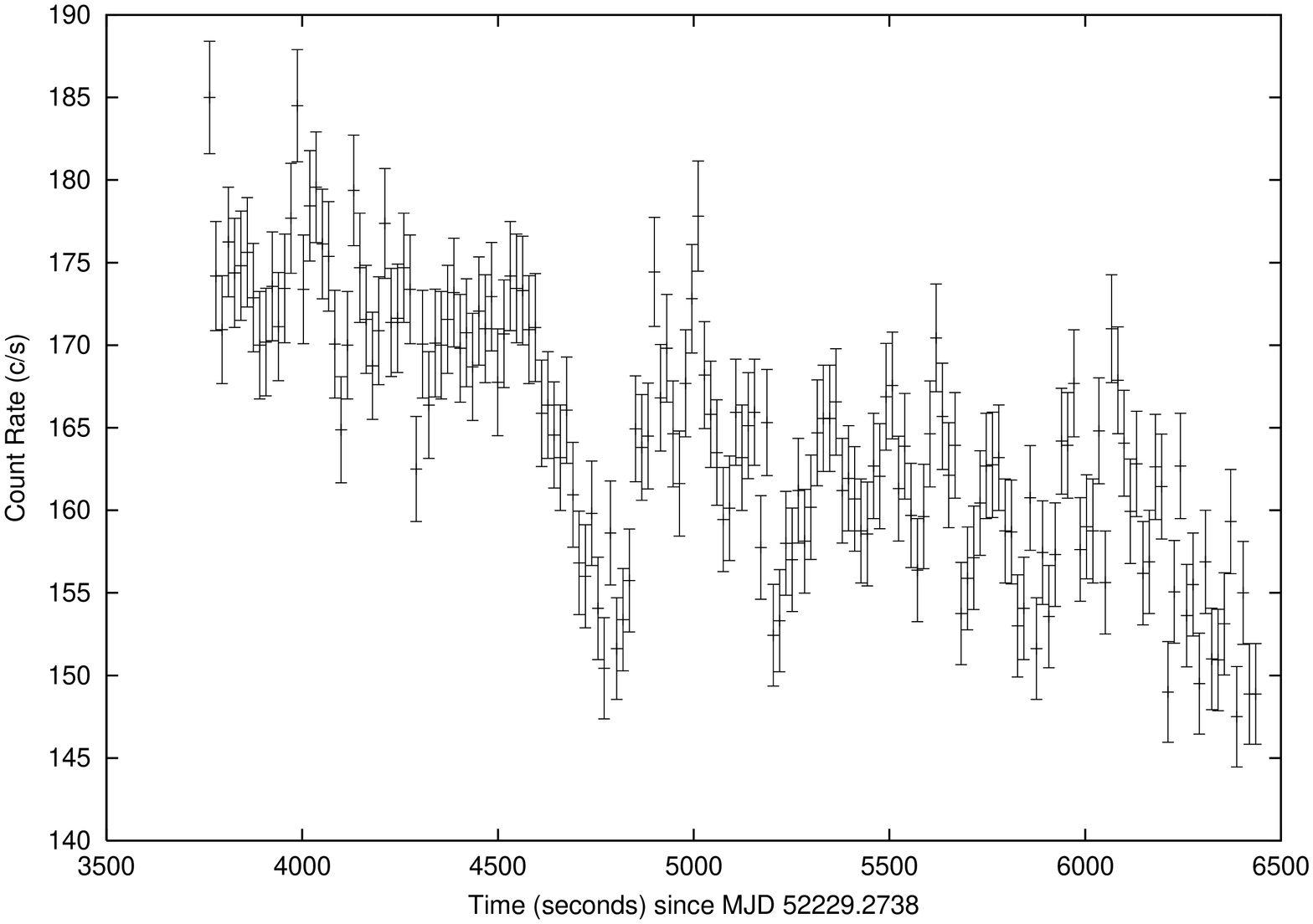,width=4.5cm, angle=270}
\end{tabular}
}
\caption{(Left): Autocorrelation
function of a ROSAT observation. The solid, dash-dotted and
dotted lines are 1$\sigma$, 2$\sigma$, and 3$\sigma$
levels for a random distribution. A periodic signal
would manifest itself as a local maximum.
 (Right): RXTE light curve of an observation
obtained in Nov. 2001.}
\end{figure}
 
\vspace{-0.5cm}
\section{Progress and the future}
We have presented the first evidence for the hard X-ray/radio
anti-correlation, and for fast time variability.
The spectral behavior 
mimics the behavior seen in other microquasars
(Mirabel \& Rodriguez 1999).
The short time scales indicate that
we are, for the first time,
probing the high-energy emission regions closer
to the compact object. 
Chakrabarti et al. (2002) show that
non-steady shocks in sub-Keplerian accretion flow provides the
basic timescale of the ejection interval.

\ss\ remains unique compared to other microquasars:
1) the ratio of its X-ray flux to the jets power
is much smaller (10$^{-4}$ compared to 0.1 for GRS~1915+105),
2) it's the only microquasar with large scale
X-ray and radio lobes
resulting from the interaction between the jets and the surrounding
medium/SNR (see however the discovery by Corbel et al., this volume,
of large scale jets from a black hole),
and 3) it's the only source confirmed to have precessing jets
with Doppler shifted emission lines
 revealing a complex system with baryonic jet
 matter.
The current picture that best explains the properties of this enigmatic
object is a binary system embedded in an expanding thick disk that is
fed by the wind from a super-Eddington accretion rate (see Gies et al. 2002).
The companion star is most likely a Wolf-Rayet star dumping some
10$^{-4}$ \msunyr into the compact object. This would
explain the large infrared flux, the low X-ray luminosity,
and the difficulty to detect any pulsations.

Several burning questions still need to be addressed with
further observations:
1) the discrepancy in the distance derived for SS~433 and W50 needs
also to be resolved to derive better estimates for the mass outflow and the
jets power; 2) the nature of the compact object remains
ambiguous. While the majority of published papers point to an underlying black hole
(e.g. Zwitter \& Calvani 1989), a later estimate
by D'Odorico \etal (1991) indicates a neutron star origin. 
Interestingly, even in the latter scenario, 
the neutron star's mass is unusually low.
If the estimated mass is indeed 0.8$\pm$0.1 \msun, then \ss\
would join the growing class of low-mass `neutron' stars
(Gondek-Rosinska, Kluzniak, \& Stergioulas 2002).

\ss\ remains unique.
While other objects have been claimed to be SS~433-like
(see reviews in this volume by Safi-Harb, Fender, and the study of
CI-Cam and V4641 Sgr by Rupen),
none of them is an \ss--twin. SNRs
offer a promising laboratory to search for
binary compact objects (potential SS~433's).
Simultaneous monitoring of \ss\ in the radio,
infrared, high energy X-rays and gamma-rays is needed to
study in more detail the spectral and timing behavior
discussed here,
in comparison with other microquasars.
Most importantly, a reliable estimate of the mass function will
hopefully solve a 25-year old puzzle.

\section*{Acknowledgments} SSH thanks NSERC for the support and the organizers for 
the invitation.

\end{document}